\documentclass[%
 reprint,
 amsmath,amssymb,
 aps,
 prb
]{revtex4-2}

\usepackage{graphicx}
\usepackage{dcolumn}
\usepackage{bm}
\usepackage{xcolor} 
\usepackage[version=4]{mhchem} 
\usepackage{graphicx}
\usepackage{xcolor}

\usepackage[linktocpage=true,
  colorlinks=true, 
  pdfborder={0 0 0}, 
  linkcolor=blue,
  citecolor=red,
  filecolor=yellow,
  urlcolor=blue,
  bookmarks,
  pdfauthor={},
]{hyperref}

\begin{document}
\title{Fragility of the magnetic order in the prototypical altermagnet \ce{RuO2}}

\author{Andriy Smolyanyuk}%
\email{andriy.smolyanyuk@tuwien.ac.at}
\affiliation{Institute of Solid State Physics, TU Wien, 1040 Vienna, Austria}
\author{Igor I. Mazin}
\email{imazin2@gmu.edu}
\affiliation{George Mason University, Department of Physics \& Astronomy and Center for Quantum Science and Engineering,  Fairfax, USA}
\author{Laura Garcia-Gassull}
\affiliation{%
Institut f\"ur Theoretische Physik, Goethe-Universit\"at Frankfurt, 60438 Frankfurt am Main, Germany
}%
\author{Roser Valent\'i}
\affiliation{%
Institut f\"ur Theoretische Physik, Goethe-Universit\"at Frankfurt, 60438 Frankfurt am Main, Germany
}%

\date{\today}

\begin{abstract}
Altermagnetism is a topic that has lately been gaining attention and the \ce{RuO2} compound is among one of the most studied altermagnetic candidates.
However, the survey of available literature on \ce{RuO2} properties suggests that there is no consensus about the magnetism of this material.
By performing density functional theory calculations, we show 
that the electronic properties of stoichiometric \ce{RuO2} are described
in terms of a smaller Hubbard $U$ within DFT+$U$ than the value required to 
have magnetism.  We further argue that \ce{Ru} vacancies can actually aid the formation of a magnetic state in \ce{RuO2}. This in turn suggests that a characterization of the amount of \ce{Ru} vacancies in experimental samples might help the resolution of the controversy between the different experimental results.

\end{abstract}

\maketitle
\section{Introduction}
In recent years the topic of altermagnetism has been gaining attention, with significant efforts directed towards finding new altermagnetic materials~\cite{PRXreview,editorial}.
Altermagnetism is defined as a magnetic phase with symmetry-driven compensated net magnetization, where the symmetry operation responsible for this magnetic phase is neither inversion nor translation. A material exhibiting these properties combines characteristics of both ferromagnetism and antiferromagnetism. Furthermore, in regards to the electronic band structure, the bands in this phase are non-spin-degenerate, leading to intriguing applications.

For a metallic system, altermagnetism would imply a possibility of generating spin-polarized currents, as in ferromagnets.
Moreover, generating spin-transfer torque and observing giant/tunneling magnetoresistance effects in such a system should be possible.
These effects are utilized to construct magnetic memory devices and the benefits that altermagnets bring in comparison to ferromagnets are the vanishing stray magnetic fields and THz range of switching~\cite{PRXreview}.

Among the various proposed materials as altermagnetic candidates, \ce{RuO2} is attracting much attention.
However, the magnetism in this system is still in itself a controversial topic.
On the one hand, the absence of a discernible phase transition in the heat capacity~\cite{cordfunke_thermophysical_1989, oneill_gibbs_1997}, magnetic susceptibility~\cite{guthrie_magnetic_1931,ryden_magnetic_1970, fletcher_magnetic_1968} and the resistivity data~\cite{lin_low_2004, glassford_electron_1994} suggests that \ce{RuO2} is a Pauli paramagnet.
On the other hand, the existence of an antiferromagnetic configuration has been reported by resonant X-ray scattering~\cite{zhu_anomalous_2019} and neutron diffraction~\cite{berlijn_itinerant_2017}. However, the latter measurements reported a rather small local magnetization value (0.05~$\mu_B$).
At the same time, no magnetic hyperfine field was detected on Ru in an NMR experiment~\cite{mukuda_spin_1999}, usually a very sensitive probe.
Additionally, there have been observations of a sizeable anomalous Hall effect, consistent with a considerably larger magnetization~\cite{feng_anomalous_2022, tschirner_saturation_2023}.

\begin{figure}[t!]
\includegraphics[width=0.8\linewidth]{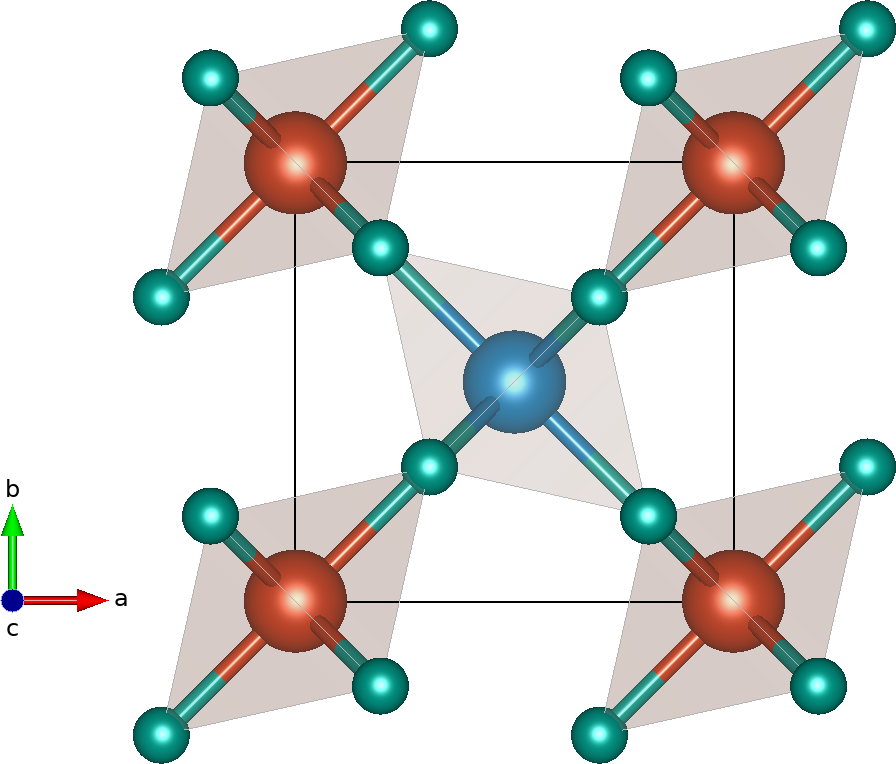}
\caption{Crystal structure of \ce{RuO2}: Ru atoms are shown in red and blue
(different colors denote different spin orientations), O atoms are shown in teal.}
\label{fig:RuO2_crystal}
\end{figure}

The magnetic configuration suggested in~\cite{zhu_anomalous_2019} is where the magnetic axis is along the $c$ axis (see Fig.~\ref{fig:RuO2_crystal}).
However, they report a better fit to the measured scattered intensity of X-rays when the magnetic moments are slightly canted.
Ref.~\onlinecite{zhu_anomalous_2019} findings and the conclusion about the long-range magnetic order are questioned by Ref.~\onlinecite{lovesey_magnetic_2022}.
In turn, they propose three different motifs (including quadrupoles into the consideration) and attempt to fit the abovementioned experimental data but have been unable to achieve a high-quality fit to any of those models.
For one of the motifs, Ref.~\onlinecite{lovesey_magnetic_2023} further provides an extended set of calculated diffraction patterns that could be used to test the presence of assumed magnetic order.
Thus, to draw an unambiguous conclusion, one would require more experimental data.

Unfortunately, the available neutron diffraction data on \ce{RuO2}~\cite{berlijn_itinerant_2017} are not sufficient to confidently resolve the controversy on the magnetization, for the reasons described below.
The main issue is that the quality of the magnetic component of the fit in these
experiments depends on the quality of the structural refinement.
In reference~\cite{berlijn_itinerant_2017}, the authors mention the possibility of a structural distortion in the rutile phase accompanied by antiferromagnetic order. However, 
Ref.~\cite{berlijn_itinerant_2017} was unable to find a distorted structure that would fit both unpolarized and polarized neutron diffraction data, while the powder X-ray diffraction patterns are consistent with the undistorted rutile structure
(see the crystal structure depicted in Fig.~\ref{fig:RuO2_crystal}).
To address this problem, Ref.~\cite{berlijn_itinerant_2017} employed density functional theory (DFT) calculations in an attempt to find such a structure. A distorted $2\times 2\times 2$ rutile supercell was optimized in both the non-magnetic and antiferromagnetic states, and the rutile structure was obtained as the ground state. 
The available computational data on the lattice dynamics in \ce{RuO2}~\cite{bohnen_lattice_2007, uchida_superconductivity_2020} confirm that the rutile structure is dynamically stable.

The absence of a structural phase transition is indirectly confirmed by other measurements. Two independent electron transport measurements, one conducted up to 300~K~\cite{lin_low_2004} and one up to 1000~K~\cite{glassford_electron_1994}, show no changes in resistivity that could be caused by a structural phase transition.
Both data sets are well described by a model that has three contributions to the resistivity: the
electron-phonon interaction with acoustic (Bloch-Gr\"unesien) and optical modes, along with a term arising from electron-electron scattering.
Moreover, there is no indication of a structural phase transition in the heat capacity measurements, which was measured up to $\sim$340~K~\cite{cordfunke_thermophysical_1989} and $\sim$1050~K~\cite{oneill_gibbs_1997}.
This same conclusion is supported by the available measurements of thermal expansion~\cite{touloukian_thermophysical_1977}.

Based on the refinement using the rutile structure, the extracted magnetic moment per \ce{Ru} atom is 0.23~$\mu_B$ for unpolarized neutron diffraction and 0.05~$\mu_B$ for polarized neutron diffraction measurements~\cite{berlijn_itinerant_2017}.
Furthermore, there is no evidence of a phase transition to an antiferromagnetic phase, neither in the susceptibility data of Ref.~\onlinecite{berlijn_itinerant_2017} nor in earlier measurements~\cite{guthrie_magnetic_1931,ryden_magnetic_1970, fletcher_magnetic_1968}.
Additionally, nuclear magnetic resonance measurements strongly suggest the absence of long-range magnetic order. This conclusion is supported by the absence of any contribution from Ru d electrons in both the Knight shift and relaxation rate, as well as the absence of any hyperfine splitting~\cite{mukuda_spin_1999}. The authors of this paper point out that, overall, the resonant magnetic properties closely resemble those of nonmagnetic Ru metal.

The controversy among the different experiments suggests that the existence of antiferromagnetic (and hence altermagnetic) order in RuO$_2$ is rather fragile, likely sample-dependent, and possibly present in only a fraction of the sample volume. In order to gain a better microscopic understanding of the magnetism (or lack thereof) in this material, we have systematically investigated the magnetic states of \ce{RuO2} employing density functional theory (DFT), both with and without a Hubbard $U$ correction applied to the Ru $d$-orbitals. Our tentative conclusion is that the perfectly ordered, stoichiometric \ce{RuO2} is likely nonmagnetic, consistent with numerous experiments above. On top of that, a modest hole doping, for instance, by creating \ce{Ru} vacancies
(a common defect in this class of materials~\footnote{Another possible source of vacancies could be an excess of oxygen, which is not unlikely since the synthesis process involves an intake of oxygen gas.
The literature reports on the stoichiometry of \ce{RuO2} are scarce and we found only one example, Ref.~\onlinecite{schafer_zur_1963}, reporting a surplus of oxygen.
However, both scenarios would lead to hole doping.
}, cf. Ref.~\onlinecite{Diulus} that found 5\% vacancies in their RuO$_2$ samples, Ref.~\onlinecite{bolzan_structural_1997} reporting 5.3\% of vacancies and Ref.~\onlinecite{rogers_crystal_1969} reporting 1\%)
promotes the \ce{RuO2} to a magnetic state of exactly the same symmetry as suggested in Ref.~\onlinecite{berlijn_itinerant_2017} and utilized in Refs.~\onlinecite{Libor_crystal_2023, Libor_chiral_magnons_2023}.

The amount of \ce{Ru} vacancies is liable to vary from sample to sample, and even from one batch to another, depending on the growth procedure, and may even be nonuniform over a sample. This could explain the discrepancy between different experiments and leads us to conclude that a characterization of the \ce{Ru} vacancies in the samples may be key to know about the magnetic character of \ce{RuO2}.

\section{Results}
\subsection{Stoichiometric \ce{RuO2}}
\begin{figure}[t!]
\includegraphics[width=1.1\linewidth]{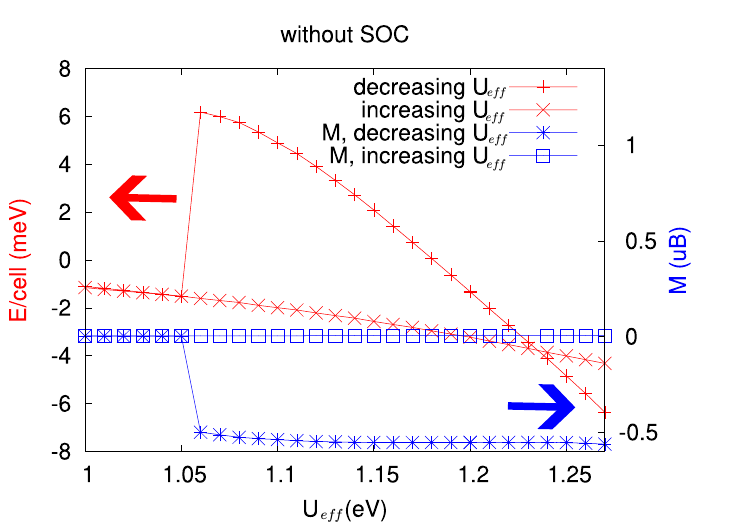}
\caption{Total energy (left) and local magnetization at the Ru site (right) as a 
function of $U_{eff}=U{-}J$ are explored in two sets of calculations. In one set (``increasing $U_{eff}$''), denoted with $\times$ (energy results) and $\Box$  symbols (magnetization results) the calculations were done starting from $U=0$~eV and progressively increasing $U$ in each subsequent calculation.
In the other set (``decreasing $U_{eff}$''), denoted with $+$ and $\ast$ symbols, the direction of the calculation was reversed. The calculations are without SOC contributions.
}
\label{fig:magn_on_U}
\end{figure}

The first concern to cover is whether stoichiometric \ce{RuO2} is magnetic or not.
To account for the possible effects of electronic correlations in this system, we perform DFT+$U$ computations.
In Fig.~\ref{fig:magn_on_U}, we plot the dependency of the local magnetic moment at the Ru site for an anti-parallel spin orientation (a parallel orientation, as well as various magnetic arrangements with $q\neq 0$, are invariably higher in energy) over a range of $U$ values.
Two sets of calculations were performed to mitigate a problem of multiple local minima inherent to DFT+$U$: one starting from $U$=0~eV and progressively increasing $U$ in each subsequent calculation and the other with decreasing $U$ (the former calculation is trapped in the non-magnetic state for the displayed range of $U$ values).
As seen from the overlap of the magnetic moment for small values of $U$ in the plot, \ce{RuO2} is non-magnetic up to a critical value, $U_{eff} =U-J {\sim}1.06$~eV.
However, in Fig.~\ref{fig:magn_on_U} only at $U_{eff} \sim 1.23$~eV the altermagnetic state becomes lower in energy than the non-magnetic: it is metastable between 1.06 and 1.23~eV.
There is a discontinuous jump (see below the explanation) of the value of the magnetic moment to ${\sim}0.5$~$\mu_B$. This jump is an order of magnitude larger than the 0.05~$\mu_B$ obtained from polarized neutron scattering measurements \cite{berlijn_itinerant_2017}, and more than twice larger than the 0.23~$\mu_B$ value fitted to unpolarized data (claimed to be less reliable and contaminated by unknown structural factors). 

\begin{figure}[t!]
\includegraphics[width=\linewidth]{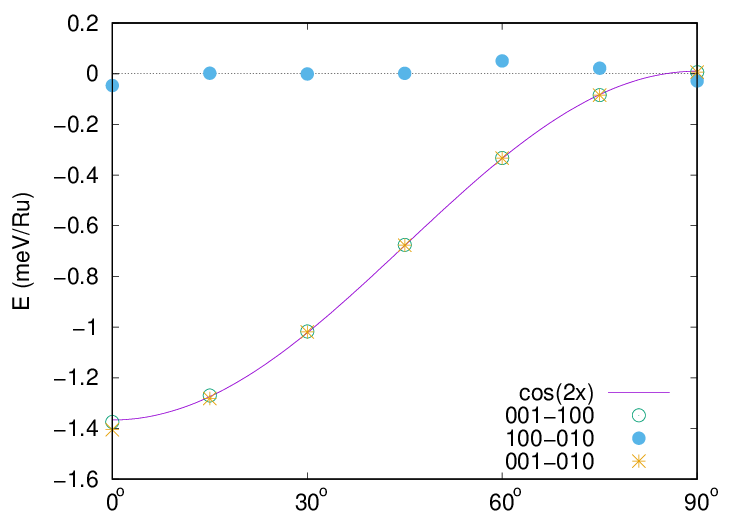}
\caption{Magnetic anisotropy as a function of the angle in degrees, when rotating the Neel vector in the indicated planes.
}
\label{fig:mae}
\end{figure}
The next issue is to focus on the magnetic ground state when the system is in the regime where magnetization is permitted.
To address this question, we initially calculated, using VASP \cite{kresse_ab_1993,kresse_ab_1994,kresse_efficiency_1996,kresse_efficient_1996}, the energy difference between the ferromagnetic (FM)
and altermagnetic (AM) stoichiometric \ce{RuO2} configurations. The calculations were done without considering SOC effects while setting the value of $U_{eff}$ to 1.3 and 1.4 ~eV and we analyzed the magnetization of both configurations.
Interestingly, the AM configuration converged to $M_{Ru}=0.66$ and 0.78~$\mu_B$ for each \(Ru\) atom in the cell, respectively. In contrast, the FM ones essentially collapsed, yielding a total magnetization of $M_{tot}=0.015$ (0.038)~$\mu_B$ per Ru atom. Correspondingly, the AM energy was lower than the FM one by 3.3 (9.3) meV/Ru.
Altogether, these results indicate that the altermagnetic configuration has the lowest energy.
Besides that, the spin-spirals with $\vec{q}=(0,0,q)$ and $\vec{q}=(q,0,0)$ were checked
leading to $q=0$ as the lowest energy state in both cases.

We also checked the calculated magnetic anisotropy and compared it with the experiment. Including spin-orbit coupling, we found that the $c$ axis is the easy axis along the 
direction, in agreement with experiment\cite{berlijn_itinerant_2017}, as seen in Fig.~\ref{fig:mae}. 

Thus, the magnetic ground state is characterized by an antiparallel alignment along the $c$ axis of the magnetic moments of two Ru atoms. This state can be described by the
magnetic space group P4$'$$_2$/mnm$'$ (BNS 136.499).

However, $U_{eff}>1$~eV is rather large for this good-metallic, strongly-hybridized,
4$d$ system.
For comparison, first principles calculations of $U_{eff}$ for the ruthenium-based spin-orbit Mott insulators $\alpha$-RuCl$_3$, RuBr$_3$, and RuI$_3$ gave estimates of 2 to 1 eV ~\cite{sibling}. Considering the metallic screening occurring in RuO$_2$, it is expected that its  $U_{eff}$  will be noticeably smaller than the values given above.
This leads us to conclude that for stoichiometric \ce{RuO2} a smaller $U_{eff}$ is likely more realistic to describe its properties than the required one to have magnetism, and therefore stoichiometric \ce{RuO2} is most probably non-magnetic.

\subsection{Density of States}
\begin{figure}[t!]
\includegraphics[width=0.95\linewidth]{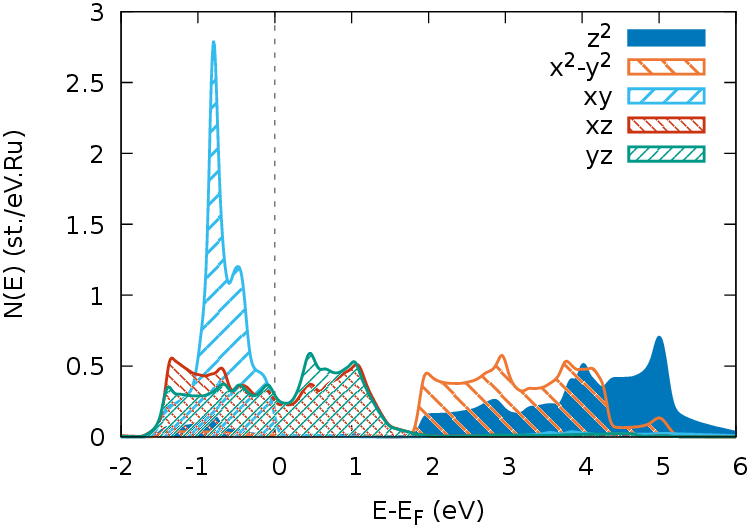}
\caption{Projected non-magnetic density of states onto \ce{Ru} $d$-orbitals: 
blue, orange, cyan, red, and teal are used to depict
$z^2$, $x^2{-}y^2$, $xy$, $xz$ and $yz$ orbitals respectively.
The coordinate system is aligned with Ru-O bonds.}
\label{fig:DOS}
\end{figure}

Moving on to analyzing the projected density of states (DOS) (see Fig.~\ref{fig:DOS}), we observe that the main contribution to the DOS around the Fermi level comes from the $xz$/$yz$ \ce{Ru} $d$-orbitals.
The value of the DOS at the Fermi level is relatively low and flat in its vicinity.
This causes the Stoner criterion for ferromagnetism to be very hard to fulfill.

Zone-center antiferromagnetism, as in RuO$_2$, obeys a modified Stoner criterion, where instead of the uniform susceptibility, $\chi(\mathbf{q}=0)=N(0)$  at some finite reciprocal lattice vectors appears,  $\chi(\mathbf{G}\ne 0)$, but, it is quite obvious that highly dispersive bands at the Fermi level and low DOS are rather unfavorable in this case as well.

Another interesting aspect in the DOS is the narrow peak below the Fermi level coming from the $xy$ \ce{Ru} orbitals.
If the Fermi level were shifted closer to this peak, it could potentially trigger a magnetic transition from the current non-magnetic state to a state with a significant magnetic moment.

Overall, the electronic structure presented here is consistent with previously reported studies~\cite{berlijn_itinerant_2017, ahn_antiferromagnetism_2019}.

\begin{figure}
\includegraphics[width=0.95\linewidth]{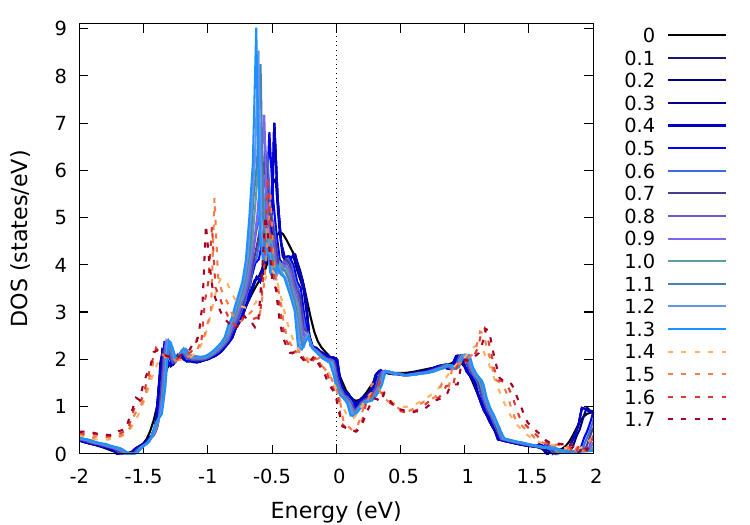}
\caption{Spin-polarized total density of states for majority spin as a function of $U_{eff}$ (in eV).
The color corresponds to various values of $U_{eff}$, where solid lines in blue shades correspond to the non-magnetic ground state and dashed lines in red shades to the AM order.
Note that here, the calculations were done using Wien2k; thus, the value of $U_{eff}$ is not directly comparable with other calculations where VASP was employed due to the differences in implementations.
}
\label{fig:TDOS}
\end{figure}

In Figure~\ref{fig:TDOS} we present the dependency of the total density of states for spin-polarized calculation on the value of $U_{eff}$.
For high values of $U_{eff}$ the $xy$ band is split but not polarized. 
That is to say, magnetic instability here is achieved not due to increased DOS, as in the below-discussed case of hole doping, but due to enhancement of the effective Stoner factor.
Indeed, as discussed in Ref.~\onlinecite{Petukhov}, in the first approximation DFT+$U$ (in its SIC flavor, used in this and other papers) is equivalent to renormalizing the DFT Stoner parameter as
$I_{eff} = I_0 + \frac{U-J}{n}$, where $I_0$ is the DFT Stoner parameter and $n$ is number of orbitals at the Fermi level~\cite{Petukhov}.
For the 4d metals, typically, $I_0\sim 0.5$ eV. 
The density of states for \ce{RuO2} at the Fermi level is $D_\uparrow(E_f)=0.7$~(st/eV$\cdot$spin), thus $D_\uparrow({Ef}) \cdot I_0 = 0.35 < 1$, so the material is stable against the formation of a ferromagnetic state.
As a rough estimate of the effect of $U$, we take $n=1$, and for $U_{eff}=1.4$ eV we get $I_{eff}\approx 1.9$~eV, thus fulfilling the Stoner criterion; for $n=2$ it is just a bit below instability.
Of course, these are just order-of-magnitude estimates, since in reality, we consider instability against antiferromagnetism, not ferromagnetism, and should instead of $D(E_F)$ use the (unknown) unrenormalized spin susceptibility at the corresponding wave vector, but it gives us the general feeling of how DFT+$U$ generate magnetism without using the $xy$ orbital.

\subsection{Non-stoichiometric \ce{RuO2}}
\begin{figure}[t!]
\includegraphics[width=\linewidth]{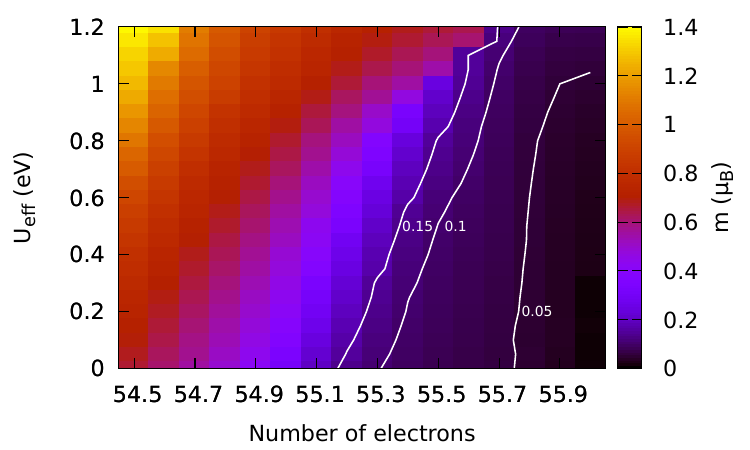}
\caption{Dependence of the local atomic magnetization $m$ on hole doping (an undoped case corresponds to 56 electrons per cell, i.e., per two formula units) and the effective Hubbard parameter $U_{eff}=U{-}J$.
Isolines for $m=0.05$, $0.10$, and $0.15$~$\mu_B$ are depicted by white lines (with no attempts to smooth the plotted lines).
}
\label{fig:magn_on_doping}
\end{figure}

The shift of the Fermi level can be accomplished through hole doping, changing the electron occupation.
The only other alternative for generating a magnetic order {\it without doping} is to increase $U$, and thus the effective Stoner parameter $I_{eff}=I+(U-J)/5$ (see Ref.~\onlinecite{Petukhov}) until the Stoner criterion ($I_{eff}>J$) is satisfied.
To check this hypothesis we did a series of DFT+$U$ calculations varying both the value of $U_{eff}$ and the number of electrons.
Figure~\ref{fig:magn_on_doping} shows the value of the local magnetic moment at the Ru site as a function of $U_{eff}$ and the number of electrons per unit cell (two formula units). 
The isoline with $m=0.05$~$\mu_B$, corresponding to the measured value from Ref.~\onlinecite{berlijn_itinerant_2017}, is highlighted. One can see that there is a rather stable ground state for $\sim 0.1$ hole/Ru doping, within a reasonable range of $U_{eff}\alt 1$ eV. For the same $U_{eff}$, a larger doping of 0.4 hole/Ru, corresponding to 10\% of Ru vacancies, generated a local magnetic moment of $m=0.2$~$\mu_B$. It is worth pointing out that the discontinuous jump in the calculated magnetic moment for the undoped compound as a function of $U_{eff}$ is immediately understood from the DOS on Fig. \ref{fig:DOS}. That is because, in order to get a stable magnetic solution, the exchange splitting (proportional to $U_{eff}$) must reach the threshold (around 1 eV, from Fig. \ref{fig:DOS}) corresponding to the separation between the $xy$ band and the Fermi level.

Note that the data in Fig.~\ref{fig:magn_on_doping} merely show a trend of the system to attain a magnetic moment with hole doping, but the preferred magnetic orientation may depend on the doping as well. 
In particular, when spin-orbit coupling (SOC) is accounted for, our DFT+$U$ calculations with $U_{eff}$=1.4~eV show that the magnetic anisotropy energy $E_{100}-E_{001}$ changes linearly with doping and there is a transition from the easy axis along the $c$ direction towards an easy plane at around 0.2 hole/Ru.
However, the full sampling of magnetic ground states as a function of $U_{eff}$ and hole doping is out of the scope of this work.

\section{Discussion}
As presented in the previous section, pristine DFT calculation characterizes stoichiometric \ce{RuO2} as non-magnetic.
Accounting for correlations with DFT+$U$ one can stabilize a non-magnetic or magnetic state depending on the value of $U_{eff}$, with a magnetic state stabilized for $U_{eff} > 1.06$~eV.
However, the stabilized magnetic state for high $U$ values is not without an issue: the value of the local magnetic moment is way higher (${\sim}0.5$~$\mu_B$ vs $0.05$~$\mu_B$) if compared with the only available value from neutron scattering measurements~\cite{berlijn_itinerant_2017}.
The report on DFT+DMFT study of this compound predicts larger magnetic moments than DFT+$U$ calculations~\cite{ahn_antiferromagnetism_2019}.
Besides that, we argue that the reasonable range of $U_{eff}$ values is when $U_{eff} < 1$~eV, leading us to conclude that stoichiometric \ce{RuO2} is non-magnetic.

Unfortunately, the available pool of experimental data does not provide a unanimous answer about \ce{RuO2} being magnetic or not.
Clearly, more measurements are needed.
For example, multiple experiments claim that \ce{RuO2} is antiferromagnetic~\cite{berlijn_itinerant_2017, zhu_anomalous_2019}, but the transition temperature has never been detected.
The temperature dependence of the local magnetic moment, if any, is unknown as well.
Furthermore, some experiments detect small but finite magnetic moment, some no long-range magnetic order, and some indirectly imply a large magnetic moment of the order of 0.6--0.7~$\mu_B$.

In this manuscript, we want to bring the attention of the scientific community to another missing piece of the puzzle: a characterization of \ce{RuO2} stoichiometry.
Our DFT calculations show that hole-doped \ce{RuO2} can be magnetic.
This finding, if confirmed experimentally, may be used to reconcile contradicting measurements: they are simply measuring different objects.

\section{Conclusions}
Our DFT calculations show that, for a realistic value of $U_{eff}$ (using \ce{Ru} based insulators as reference), the stoichiometric \ce{RuO2} compound is non-magnetic. However, hole doping due to \ce{Ru} vacancies can induce a phase transition to the antiferromagnetic phase even for small values of $U_{eff}$. This observation may be a key to reconcile different, strongly mutually contradicting experiments. 
If our conjecture is correct, every experimental work on \ce{RuO2} must begin with a careful characterization of the O and Ru content. Moreover, a systematic experimental investigation of magnetic properties as a function of the O and Ru content is absolutely necessary. One verifiable corollary is that with controllable Ru vacancies, one should be able to observe the antiferromagnetic transition in thermodynamics, transport and magnetometry.

\section{Computational Details}
Computations were done using density functional theory (DFT) in the generalized
gradient approximation (GGA) with the 
Perdew-Burke-Ernzerhof~\cite{perdew_generalized_1996,perdew_generalized_1997} functional as 
implemented in the \texttt{VASP}~\cite{kresse_ab_1993,kresse_ab_1994,kresse_efficiency_1996,kresse_efficient_1996}
package employing the projector augmented wave method (PAW)~\cite{blochl_projector_1994,kresse_ultrasoft_1999},
Ru\_sv and O (or O\_h, for structural optimization) pseudopotentials were used.
The energy cutoff was set to 400~eV (for the test purpose, selected calculations were performed with a 900 eV cutoff) and the 8x8x12 (12x12x18 for testing) k-points Monkhorst-Pack
grid~\cite{monkhorst_special_1976,pack_special_1977} was used.
The hole doping was achieved by varying the number of electrons (NELECT flag in \texttt{VASP}) and the calculations were done using a pristine rutile structure with $a=4.480$~\AA, $c=3.105$~\AA, Ru at $2a$ Wyckoff position and O at $4f$ position with $x=0.30479$~\cite{sd_0305751}.
For selected calculations, a cross-check using Wien2k was used.
Wien2k was also used to compute the total and projected density of states.

The data needed to reproduce and verify the results presented in this manuscript is publicly available on the TU Wien Research Data repository~\cite{RuO2_data}.

\section{Acknowledgments}
We thank Libor \v{S}mejkal, Huibo Cao, Dmitry Khalyavin, Jan Kune\v{s}, and Robert Svagera for the discussions.
A.S. was supported by the Austrian Science Fund (FWF) through the project P33571 ``BandITT''.
L.G-G. and R.V. were supported by the Deutsche Forschungsgemeinschaft (DFG, German
Research Foundation) for funding through TRR 288 –
422213477 (project B05). I.M. was supported by the Army Research
Office under Cooperative Agreement Number W911NF-
22-2-0173. He also acknowledges Heraeus Foundation for supporting his visits to University of Frankfurt.
Some of the images in the paper were created using VESTA software~\cite{VESTA}.

\bibliography{RuO2_paper}

\end{document}